# Normal black holes in bulge-less galaxies: the largely quiescent, merger-free growth of black holes over cosmic time

G. Martin,[1]⋆ S. Kaviraj,[1] M. Volonteri,[2] B. D. Simmons,[3]† J. E. G. Devriendt,[4]
C. J. Lintott,[4] R. J. Smethurst,[5] Y. Dubois,[2] and C. Pichon[2,6]

[1]*Centre for Astrophysics Research, School of Physics, Astronomy and Mathematics, University of Hertfordshire, College Lane, Hatfield AL10 9AB, UK*
[2]*Institut d'Astrophysique de Paris, Sorbonne Universités, UMPC Univ Paris 06 et CNRS, UMP 7095, 98 bis bd Arago, 75014 Paris, France*
[3]*Center for Astrophysics and Space Sciences (CASS), Department of Physics, University of California, San Diego, CA 92093, USA*
[4]*Dept of Physics, University of Oxford, Keble Road, Oxford OX1 3RH UK*
[5]*School of Physics and Astronomy, The University of Nottingham, University Park, Nottingham, NG7 2RD, UK*
[6]*Korea Institute for Advanced Study (KIAS), 85 Hoegiro, Dongdaemun-gu, Seoul, 02455, Republic of Korea*

29 January 2018

**ABSTRACT**

Understanding the processes that drive the formation of black holes (BHs) is a key topic in observational cosmology. While the observed $M_{\rm BH}$–$M_{\rm Bulge}$ correlation in bulge-dominated galaxies is thought to be produced by major mergers, the existence of a $M_{\rm BH}$–$M_\star$ relation, across all galaxy morphological types, suggests that BHs may be largely built by secular processes. Recent evidence that bulge-less galaxies, which are unlikely to have had significant mergers, are offset from the $M_{\rm BH}$–$M_{\rm Bulge}$ relation, but lie on the $M_{\rm BH}$–$M_\star$ relation, has strengthened this hypothesis. Nevertheless, the small size and heterogeneity of current datasets, coupled with the difficulty in measuring precise BH masses, makes it challenging to address this issue using empirical studies alone. Here, we use Horizon-AGN, a cosmological hydrodynamical simulation to probe the role of mergers in BH growth over cosmic time. We show that (1) as suggested by observations, simulated bulge-less galaxies lie offset from the main $M_{\rm BH}$–$M_{\rm Bulge}$ relation, but on the $M_{\rm BH}$–$M_\star$ relation, (2) the positions of galaxies on the $M_{\rm BH}$–$M_\star$ relation are not affected by their merger histories and (3) only ∼35 per cent of the BH mass in today's massive galaxies is directly attributable to merging – the majority (∼65 per cent) of BH growth, therefore, takes place gradually, via secular processes, over cosmic time.

## 1 INTRODUCTION

The co-evolution of galaxies and their black holes (BHs) is a central theme of our galaxy formation paradigm. In the nearby Universe, several correlations are observed between BH mass and the properties of the host galaxy, such as its velocity dispersion (Magorrian et al. 1998; Ferrarese & Merritt 2000), the mass of its bulge (e.g. Marconi & Hunt 2003; Häring & Rix 2004) and its total stellar mass (e.g. Cisternas et al. 2011a; Marleau et al. 2013), which suggest that the evolution of galaxies and their central BHs may be linked.

However, the processes that underpin these correlations have remained a matter of debate. For example, the correlation between BH and bulge mass is often considered to be a product of galaxy mergers (e.g. Sanders et al. 1988; Croton et al. 2006; Hopkins et al. 2006). Simulations show that mergers (in particular 'major' mergers, i.e those with near-equal mass ratios) are efficient at building bulges (e.g. Toomre & Toomre 1972; Barnes 1992), although some bulges may form via other processes, such as disk instabilities (e.g. Dekel et al. 2009; Kaviraj et al. 2013a) and, in cases where gas fractions are particularly high, disks may reform from residual gas even after a major merger (see e.g. Springel & Hernquist 2005; Kannappan et al. 2009; Hopkins et al. 2009). Combined with the fact that active galactic nuclei (AGN), and thus growing BHs, are often observed in systems undergoing major mergers (e.g. Urrutia et al. 2008; Bessiere et al. 2014; Chiaberge et al. 2015; Glikman et al. 2015; Trakhtenbrot et al. 2017), it is reasonable to suggest that this process could create the observed $M_{\rm BH}$–$M_{\rm Bulge}$ correlation, by simultaneously building the BH and the galaxy bulge (e.g. Sanders et al. 1988; Hopkins et al. 2006; Peng 2007).

While much of the past literature on BH–galaxy correlations has focussed on early-type (i.e. bulge dominated) galaxies, recent work has started to probe how these correlations may behave in the general galaxy population. Many studies now indicate that a broad correlation exists across the general population of galaxies, if the relationship between BH mass and the *total* stellar mass (e.g. Grier et al. 2011; Cisternas et al. 2011a,b; Marleau et al. 2013; Reines & Volonteri 2015) or the relationship between BH mass and halo mass (e.g. Booth & Schaye 2010; McAlpine et al. 2017) of the host galaxy is considered. The $M_{\rm BH}$–$M_{\rm Bulge}$ correlation is then likely to be just a subset of this general trend, since early-type galaxies are bulge-dominated, and therefore their total stellar mass is largely the same as their bulge mass.

⋆ E-mail: g.martin4@herts.ac.uk
† Einstein Fellow

The origin of a $M_{\rm BH}$–$M_\star$ correlation, that exists irrespective of morphological type, is difficult to explain via major mergers alone, since the bulk of the material contained in galaxy disks is likely to have been formed via secular processes (e.g. Martig et al. 2012; Conselice et al. 2013). However, the recent observational literature suggests that building up such a correlation, via processes other than major mergers, is plausible. AGN, particularly those with moderate accretion rates typical of normal galaxies (Hasinger et al. 2005), are often found in systems that are not associated with major mergers (e.g. Grogin et al. 2005; Gabor et al. 2009; Pipino et al. 2009; Kaviraj et al. 2012; Kocevski et al. 2012; Shabala et al. 2012; Kaviraj 2014a; Kaviraj et al. 2015a). Recent studies have shown that minor mergers can enhance star-formation and nuclear-accretion rates (e.g. Kaviraj 2014a,b; Comerford et al. 2015; Capelo et al. 2015; Smethurst et al. 2015; Steinborn et al. 2016; Martin et al. 2017) and could, therefore, produce BH growth while leaving the disk intact. Certain secular processes which are connected to, or responsible for, fuelling star formation – e.g. bar driven inflows of gas (Regan & Teuben 2004; Lin et al. 2013), disc instabilities (Bournaud et al. 2011) or cosmological cold flows (Feng et al. 2014) – may also be capable feeding the BH by driving gas towards the central regions of galaxies.

There is evidence that spiral galaxies with low central velocity dispersions, and therefore low bulge masses, tend to have overmassive BHs (Sarzi et al. 2002; Beifiori et al. 2009) when considering the $M_{\rm BH}$–$M_{\rm Bulge}$ correlation, which suggests that the processes that build the BH and the bulge may be different (e.g Grupe & Mathur 2004; Mathur & Grupe 2004, 2005a,b). It is also worth noting that a general dearth of major mergers in the AGN population is found around the epoch of peak cosmic star formation (e.g. Simmons et al. 2011; Schawinski et al. 2012; Kocevski et al. 2012), when the bulk of the stellar and BH mass in today's galaxies was assembled (e.g. Hopkins et al. 2006).

Furthermore, recent work (Martin et al. 2017) has shown that a majority (∼90 percent) of the stellar mass in today's Universe is likely to be unrelated to major merging. If BH and stellar mass growth move in lockstep with each other, then it is reasonable to suggest that the BH accretion rate budget may also be decoupled from the major-merger process, which would then lead to the $M_{\rm BH}$–$M_\star$ relation observed at low redshift.

Some caveats to the arguments presented above are worth considering. Given that BHs comprise, on average, only ∼0.2 per cent of their host galaxy's stellar mass (e.g. Häring & Rix 2004), it is still possible that they form preferentially in major mergers, since this process is responsible for ∼10 per cent of cosmic stellar growth (Martin et al. 2017). In addition, since major mergers with high gas fractions may result in reformed disks, it may be possible for 'normal' BHs to form in systems that do not have early-type morphology, but which have had gas-rich major mergers in their formation history. Additionally, recent work (e.g. Sani et al. 2011; Mathur et al. 2012; Kormendy & Ho 2013) have found that galaxies with pseudo-bulges – which are often interpreted to be the result of minor mergers and secular processes (e.g. Kormendy & Kennicutt 2004) – lie below the $M_{\rm BH}$–$M_{\rm Bulge}$ relation, suggesting that another process, such as major merging, may still be an important channel for BH growth (although it is worth noting that BH masses in pseudo-bulged galaxies do not differ greatly from those in galaxies that exhibit classical bulges).

A compelling counter to the hypothesis of (major) merger-driven black hole growth is the presence of massive black holes in disc-dominated galaxies. Studies of the BHs in such systems, (e.g Filippenko & Ho 2003; Ghosh et al. 2008; Araya Salvo et al. 2012) show that supermassive BHs cannot be associated exclusively with bulges. A particularly stringent test of whether major mergers preferentially build BHs is to compare the BH masses in bulge-less galaxies (i.e. those that are unlikely to have had many major mergers) to those in the general galaxy population. While such galaxies are rare in the nearby Universe, Simmons et al. (2017) have recently performed such a test on disk-dominated and bulge-less systems drawn from the SDSS. Their study shows that disk-dominated and bulge-less galaxies lie offset above the main locus in the $M_{\rm BH}$–$M_{\rm Bulge}$ correlation. However, these galaxies fall on the main locus of the $M_{\rm BH}$–$M_\star$ relation, like the rest of the galaxy population. In other words, bulge-less galaxies appear to have normal BHs, yet are systems that are unlikely to have had many major mergers in their evolutionary histories. In these systems, at least, major mergers appear unlikely to have been the dominant drivers of BH growth.

While recent observational work hints at the possibility that BH growth does not require major mergers (e.g. Kaviraj 2014b), it remains challenging to address this issue via empirical work alone, at least using current surveys. Current observational datasets are heterogeneous and relatively small and the measurement of precise BH masses remains a difficult exercise. Furthermore, while major mergers are generally expected to build bulges, there is a possibility that disk rebuilding after very gas-rich major mergers (especially at high redshift) may preserve some disky structure - differentiating such systems from 'normal' disk galaxies that have evolved in the absence of major mergers is difficult observationally.

Given the observational challenges described above, an alternative approach is to appeal to a theoretical model that reproduces both the stellar mass growth and BH demographics of the galaxy population over cosmic time. In this study, we use Horizon-AGN, a cosmological hydrodynamical simulation, to probe the BH–galaxy correlations that naturally arise in the standard model. We specifically probe the evolution of bulge-less galaxies in the simulation, explore how galaxies with varying contributions of major and minor mergers in their evolutionary histories differ in their positions on these correlations, and quantify how much of the BH accretion budget is directly attributable to merging over cosmic time.

This paper is organized as follows. In Section 2, we describe the simulation employed by this study. Although the model is described in detail in Dubois et al. (2014) and Kaviraj et al. (2017), we outline the treatment of baryons and BHs, as these are relevant to the observable quantities that are being studied in this paper. In Section 3, we briefly outline the role of mergers (and major mergers in particular) in the production of bulges. In Section 4, we explore the correlations between BH, bulge and total stellar mass that are produced by the simulation, the potential role of mergers in the formation of these correlations and quantify the fraction of the BH accretion budget that is directly attributable to mergers over cosmic time. We summarize our findings in Section 5.

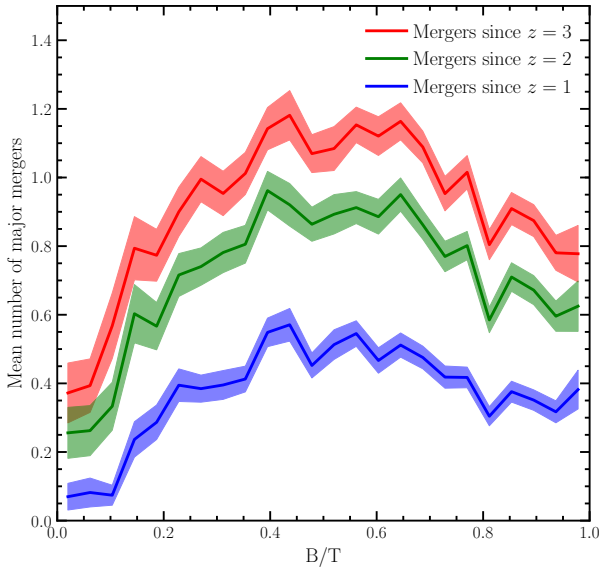

**Figure 1.** The mean number of major mergers that local massive galaxies ($M_\star > 10^{10.5}$ at $z = 0$) have undergone after a given redshift, as a function of their bulge to total stellar mass (B/T) ratios. The colour corresponding to a given redshift is indicated by the legend. Filled polygons indicate the standard error on the mean from Poisson errors. Major mergers are defined as mergers with mass ratios greater than 1 : 4. Note that the downward trend in the mean number of major mergers at high B/T values is driven by the fact that these galaxies are typically massive early-type galaxies. Since these systems tend to be some of the highest-mass systems at a given epoch, there are, by definition, not many systems of similar mass. Thus, these galaxies are less likely to experience major mergers. The number of minor mergers (not shown) is typically a factor of 2.5 times greater (e.g. Kaviraj et al. 2015a).

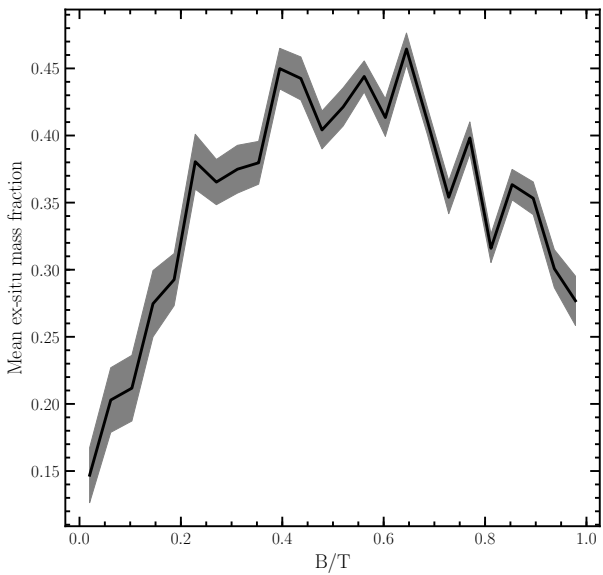

**Figure 2.** The mean fraction of ex-situ mass in today's massive galaxies ($M_\star > 10^{10.5}$ at $z = 0$), as a function of their B/T ratio. Galaxies with low B/T ratios are likely to have low ex-situ mass fractions, indicating that the majority of their stellar mass formed via secular processes.

## 2 THE HORIZON-AGN SIMULATION

Horizon-AGN is a hydrodynamical simulation (Dubois et al. 2014) in a cosmological volume that employs the adaptive mesh refinement (AMR) code, RAMSES (Teyssier 2002). The simulation box is $100 \, h^{-1}$ coMpc on a side, with $1024^3$ dark matter particles, and uses initial conditions from a *WMAP7* ΛCDM cosmology (Komatsu et al. 2011). It has a dark matter mass resolution of $8 \times 10^7 \, M_\odot$, a stellar-mass resolution of $2 \times 10^6 \, M_\odot$ and a spatial resolution of ~1 kpc. A quasi Lagrangian criterion is used to refine the initially uniform $1024^3$ grid, when 8 times the initial total matter resolution is reached in a cell, down to a minimum cell size of 1 kpc in proper units.

In the following sections, we describe some aspects of the simulation that are central to this study: the treatment of baryons, the identification of galaxies and mergers, the growth of BHs and BH feedback on ambient gas. As described in detail in Kaviraj et al. (2017), Horizon-AGN reproduces key observables that trace the aggregate cosmic stellar mass growth of galaxies since $z \sim 6$: stellar mass and luminosity functions, rest-frame UV-optical-near infrared colours, the star formation main sequence, the cosmic star formation history and galaxy merger histories (Kaviraj et al. 2015b). It also reproduces the demographics of black holes (BHs) over cosmic time: the BH luminosity and mass functions, the BH mass density versus redshift, and correlations between BH and galaxy mass (Volonteri et al. 2016).

### 2.1 Treatment of baryons

Following Sutherland & Dopita (1993), gas cools via H, He and metals, down to $10^4$ K. A UV background is switched on at $z = 10$, following Haardt & Madau (1996). Star formation takes place via a standard 2 per cent efficiency (Kennicutt 1998), when the hydrogen density reaches a critical threshold of $n_0 = 0.1$ H cm$^{-3}$. A subgrid model for stellar feedback is implemented, that includes all processes that may impart thermal and kinetic feedback on ambient gas.

Horizon-AGN implements continuous stellar feedback that incorporates momentum, mechanical energy and metals from stellar winds and Type II/Type 1a SNe. When considering stellar winds and Type II SNe, the STARBURST99 model (Leitherer et al. 1999, 2010) is employed to generate look-up tables as a function of metallicity and age. The model employs the Padova tracks (Girardi et al. 2000), with thermally pulsating asymptotic branch stars (see e.g. Vassiliadis & Wood 1993). The kinetic energy of the stellar winds is calculated using the 'Evolution' model of Leitherer et al. (1992).

The implementation of Type Ia SNe follows Matteucci & Greggio (1986) and assumes a binary fraction of 5% (Matteucci & Recchi 2001). The chemical yields are taken from the W7 model of Nomoto et al. (2007). Stellar feedback is modelled as a heat source after 50 Myr (mainly to reduce computational cost). This is reasonable, given that, after 50 Myr, the bulk of the energy from stellar feedback is liberated via Type Ia SNe that have long time delays i.e. several hundred Myrs to a few Gyrs (e.g. Maoz et al. 2012). These systems are not prone to strong radiative losses, as stars disrupt their dense birth clouds, or move away from them, after a few tens of Myrs (e.g. Blitz & Shu 1980; Hartmann et al. 2001).

### 2.2 Identification of galaxies and mergers

We identify galaxies using the ADAPTAHOP structure finder (Aubert et al. 2004; Tweed et al. 2009), which is applied to the distribution of star particles. Structures are identified using a local threshold of 178 times the average matter density. The local density of individual star particles is measured using the 20 nearest neighbours, with structures that have more than 50 particles being considered as galaxies. This corresponds to a minimum identifiable stellar mass of $10^{8.5}$ $M_\odot$ and yields a catalogue of $\sim$100,000 galaxies with $M_\star > 10^9$ $M_\odot$ at $z = 0.06$. We then produce merger trees for each galaxy, tracking their progenitors to $z = 6$. The average length of timesteps is $\sim$130 Myr.

We use the merger trees to identify the major (mass ratios $> 1:4$) and minor (mass ratios between $1:4$ and $1:10$) mergers that each galaxy has undergone. It is worth noting that how the mass ratio is defined influences the identification of major and minor mergers (e.g. Rodriguez-Gomez et al. 2015). Here, we adopt the mass ratio when the mass of the less massive galaxy is at its maximum prior to coalescence – i.e. before material is transferred between the merging companions. This effectively measures the 'true' mass ratio of the system, before the merger process alters the properties of galaxies involved in the merger.

We note that BH growth is not prescriptively linked to mergers in the simulation. The growth in BH mass is simply a result of accretion from ambient gas, but will naturally respond to changes in the geometry and dynamics of the gas that is induced by a merger e.g. if major mergers efficiently funnel gas into the centre of a remnant then BH growth could be accelerated. However, the model is not set up to preferentially build BHs during mergers.

The minimum galaxy mass of $10^{8.5}$ $M_\odot$ imposes a limit on the minimum merger mass ratio that is detectable for a galaxy of a given mass. For example, of galaxies that have stellar masses around $10^{9.5} M_\odot$ at $z = 0.06$ (the final snapshot of the simulation), 96, 72 and 37 per cent are massive enough to detect a merger with a mass ratio of $1:4$, at $z = 1$, $z = 2$ and $z = 3$ respectively. For mergers with mass ratios of $1:10$, the corresponding values are 84, 47 and 20 per cent at the same redshifts. For galaxies with stellar masses above $10^{11} M_\odot$, the merger history is at least 85 per cent complete for mass ratios greater than $1:10$, up to $z = 3$.

### 2.3 The growth of black holes and black-hole feedback on ambient gas

BH or 'sink' particles are seeded in the simulation wherever the gas density in a cell exceeds a critical threshold of $\rho > \rho_0$ and the stellar velocity dispersion exceeds 100 km s$^{-1}$, where $\rho_0 = 1.67 \times 10^{-25}$ g cm$^{-3}$ and corresponds to 0.1 H cm$^{-3}$, the minimum threshold for star formation. To prevent the formation of multiple BHs within the same galaxy, BHs cannot form while there is another BH within 50 kpc. BHs have an initial mass of $10^5$ $M_\odot$, which is chosen to match BH masses predicted a direct collapse scenario (e.g. Begelman et al. 2006). However, BH masses quickly become self regulated, so that the exact choice of seed mass is not important (Dubois et al. 2012).

BH seeding continues until $z = 1.5$, after which no new BHs are allowed to form. This is purely to prevent an unmanageable number of BHs from being formed, and has a negligible effect on the the growth of massive BHs. Almost all late forming BHs do so in low mass galaxies, and by $z = 0$, the BH occupation fractions of massive galaxies are in agreement with observational estimates (e.g. Trump et al. 2015).

Following their formation, each BH is able to grow through gas accretion, or through coalescence with another black hole (Dubois et al. 2014, 2016). Accretion is modelled using the Bondi-Hoyle-Lyttleton rate:

$$\dot{M}_{BH} = \frac{4\pi\alpha G^2 M_{\rm BH}^2 \bar{\rho}}{(\bar{c}_s^2 + \bar{u}^2)^{3/2}}, \quad (1)$$

where $M_{\rm BH}$ is the mass of the BH, $\bar{\rho}$ is the mass-weighted average gas density, $\bar{c}_s$ is the mass-weighted average sound speed, $\bar{u}$ is the mass-weighted average gas velocity relative to the BH and $\alpha$ is a dimensionless boost factor which accounts for the inability of the simulation to capture the cold high-density inter-stellar medium and corrects for accretion that is missed due to unresolved gas properties (Booth & Schaye 2009). The effective accretion rate of the BH is capped at the Eddington accretion rate:

$$\dot{M}_{\rm Edd} = \frac{4\pi G M_{\rm BH} m_p}{\varepsilon_r \sigma_T c}, \quad (2)$$

where $m_p$ is the mass of a proton, $\varepsilon_r$ is the radiative efficiency, assumed to be $\varepsilon_r = 0.1$ for Shakura & Sunyaev (1973) accretion onto a Schwarzschild BH, $\sigma_T$ is the Thompson cross-section and $c$ is the speed of light. BHs are allowed to coalesce if they form a tight enough binary. Two black holes must be within four AMR cells of one another and have a relative velocity that is smaller than the escape velocity of the binary. The resulting mass of the merged binary is simply the sum of the masses of the two BHs.

BH feedback on ambient gas operates via a combination of two channels and depends on the ratio of the gas accretion rate to the Eddington luminosity, $\chi = \dot{M}_{BH}/\dot{M}_{\rm Edd}$. For Eddington ratios $\chi > 0.01$ (which represent high accretion rates) a 'quasar' mode is implemented, with 1.5 per cent of the accretion energy being injected isotropically into the gas as thermal energy. For Eddington ratios $\chi < 0.01$ (which represent low accretion rates) a 'radio' mode is active, where cylindrical bipolar outflows are implemented with a jet velocity of $10^4$ km s$^{-1}$. The quasar mode efficiency is chosen to reproduce the local $M_{\rm BH}$–$M_\star$ and $M_{\rm BH}$–$\sigma_\star$ relations, as well as the local cosmic black-hole mass density (Dubois et al. 2012). Horizon-AGN is not otherwise tuned to reproduce the bulk observable properties of galaxies in the nearby Universe.

The effect of AGN feedback in Horizon-AGN is to regulate BH growth and star formation by preventing the accumulation of cold gas (Dubois et al. 2012, 2016). Rapid cosmological accretion in the early Universe leads to enhanced quasar mode activity and is the dominant mode of feedback for high redshift, gas-rich galaxies. As galaxies grow, they expel or consume their supply of cold gas leading to reduced BH accretion rates. As a result, the radio mode becomes increasingly important towards lower redshifts, eventually becoming the dominant mode of feedback in the low redshift Universe (Krongold et al. 2007; Best & Heckman 2012; Dubois et al. 2012; Volonteri et al. 2016; Peirani et al. 2017).

Rather than being anchored to the centre of their dark matter haloes as in some other simulations (e.g Taylor & Kobayashi 2014; Schaye et al. 2015; Sijacki et al. 2015), BHs are allowed to move freely, with a drag force applied in order to mitigate unrealistic motions and spurious oscillations arising from the effect of a finite particle resolution. BHs must, therefore, be matched with a host galaxy, since they are not explicitly assigned to a host galaxy by the simulation. We assign a BH to a host galaxy only if it lies within twice the effective radius of a galaxy structure and within 10 per cent of the virial radius of its dark matter halo. By this definition, a majority of massive galaxies ($M_\star > 10^{10} M_\odot$) at $z \sim 0$ are host to a BH (Volonteri et al. 2016). In practice, almost all single luminous ($L_{bol} > 10^{43}$) BHs are found at the centres of their host galaxies.

Binary BHs account for a significant fraction of the off-centre BH population, with single off-centre BHs accounting for less than 1 per cent of the total population of luminous BHs (Volonteri et al. 2016).

## 2.4 Galaxy morphology: measurement of B/T ratios

We employ bulge-to-total (B/T) ratios calculated by Volonteri et al. (2016). Sérsic fits to the stellar mass profiles of our simulated galaxies are performed, which include a disc component with index $n = 1$, plus a second 'bulge' component with index $n = [1, 2, 3, 4]$, with the best-fitting component used for our analysis. The mass associated with each component is measured, from which the B/T ratio is calculated.

Observational studies of bulge-less galaxies can differ slightly in their definition of a 'bulge-less' system (e.g. Kormendy et al. 2010; Jiang et al. 2011; Simmons et al. 2012, 2013; Secrest et al. 2012; Marleau et al. 2013). Here, we follow Marleau, Clancy & Bianconi (2013) and define objects as bulge-less if they have B/T $< 0.1$. In the final snapshot of the simulation ($z = 0.06$), 2.8 per cent of galaxies are classified as bulge-less. At $z = 0.5$ and $z = 2.5$, the corresponding values are 2.5 per cent and 5.5 per cent respectively. Note that the second component of the fit in around half of these bulge-less galaxies has an index of $n = 2$ or below, (with a similar fraction for galaxies that are not classified as bulge-less), possibly indicating a pseudo-bulge, in broad agreement with Simard et al. (2011) for sufficiently resolved galaxies.

## 3 MERGERS AND PRODUCTION OF BULGES

The role of galaxy mergers in driving morphological transformation, as a function of the properties of the merging progenitors (e.g. stellar mass, gas fraction, orbital configuration, local environment and redshift) will be addressed in detail in a forthcoming paper (Martin et al. in prep). Here, we outline some aspects of merger-driven bulge formation that are relevant to this study.

We begin by exploring the hypothesis that mergers are primarily responsible for the production of bulges and, as is often assumed in observational studies (e.g. Kormendy & Kennicutt 2004; Satyapal et al. 2009, 2014; Schawinski et al. 2011; Bizzocchi et al. 2014), that galaxies that do not contain significant bulge components (e.g. those with B/T ratios less than 0.1) must not have undergone significant merger activity. This assumption is typically motivated by idealised simulations of isolated mergers (e.g. Di Matteo et al. 2005; Hopkins et al. 2006) which do not, therefore, place the merging system in a cosmological context or realistically sample the parameter space. While they share similar physics to their cosmological counterparts, e.g. in terms of prescriptions for BH growth (typically Eddington-limited Bondi accretion), star formation and implementation of other baryonic processes, idealised simulations do not allow for statistical studies of galaxy evolution, nor do they model a galaxy's wider environment. They cannot, for example, account for cosmological accretion from filaments or cooling of hot halo gas, which may contribute to continued stellar mass growth and rebuilding of disks subsequent to the merger.

Disks could also regrow simply from residual gas from the merger progenitors, in cases where the initial gas fractions are extremely high. Such processes could act to increase the total mass of the galaxy, without necessarily growing the bulge, and therefore work to reduce the B/T ratio of the galaxy. The assumption that bulge-less galaxies have undergone no major mergers could, therefore, depend somewhat on the epoch at which a merger takes place, and the accretion and star formation history of the galaxy (e.g Sparre & Springel 2017). For example, galaxies in the high-redshift Universe exhibit high gas fractions (e.g. Tacconi et al. 2010; Geach et al. 2011), which may enable (gas-rich) merger remnants to re-grow disks, either via cosmological accretion and/or gas left over after the merger (e.g. Springel & Hernquist 2005; Athanassoula et al. 2016; Font et al. 2017).

In Figure 1, we study the effect of major mergers (mass ratios > 1 : 4) on the B/T fraction of galaxies in the local Universe. For the purposes of the analysis in this section, we limit ourselves to galaxies with stellar masses greater than $10^{10.5} M_\odot$, because their merger histories are relatively complete. Our sample is > 80 per cent complete for mergers of mass ratios 1 : 4 at $z = 3$ (i.e. more than 80 per cent of $10^{10.5} M_\odot$ galaxies were massive enough at $z = 3$ that a 1 : 4 merger would be detectable) and > 95 per cent complete at $z = 1$ and $z = 2$.

Figure 1 shows the mean number of major mergers as a function of the final B/T ratio of the local galaxy. Blue, green and red lines indicate the mean number of mergers since $z = 1$, $z = 2$ and $z = 3$ respectively. The probability that bulge-less galaxies (B/T < 0.1) have undergone any major mergers between $z = 1$ and the present day is essentially zero. However, while the assumption that bulge-less galaxies have undergone no major mergers after $z = 1$ appears to be a good one, it is necessary to relax this somewhat towards higher redshifts. For example, bulge-less galaxies have, on average, ∼0.25 and ∼ 0.35 major mergers since $z = 2$ and $z = 3$ respectively. In other words, around one in four and one in three of these galaxies have undergone a major merger since $z = 2$ and $z = 3$ respectively. Note that a qualitatively similar picture emerges when considering minor mergers (mass ratios between 1 : 4 and 1 : 10). A small fraction (∼ 20 per cent) of galaxies with low B/T values have had a minor merger since $z = 1$, indicating that some galaxies have survived recent low mass ratio mergers, without producing a significant bulge component.

The major-merger histories described above suggest that disk rebuilding plays some role in the merger history of even those galaxies that do not exhibit a strong bulge component at the present day (although such events are relatively rare). The effect of disk rebuilding (e.g. from cosmological accretion of cold gas or continued stellar mass growth from residual gas after a merger) is largely to wash out some of the morphological (i.e. disk to bulge) transformation produced by high-redshift mergers. We note that, for high values of B/T ($> 0.7$), the number of major mergers decreases. This is because the most massive galaxies (e.g. $M_* > 10^{11} M_\odot$) typically dominate the high B/T population (e.g. Dubois et al. 2016) and there are, by definition, not many systems of similar mass. Thus, these galaxies are less likely to experience major mergers.

It is also instructive to directly consider the fraction of stellar mass in a galaxy that did not form in-situ. Using a raw number of mergers could be misleading, because the impact of a merger on the final morphology of a galaxy at $z = 0$ depends, to some extent, on the final mass of the galaxy, in addition to the mass of the galaxy at the time of the merger, since, as discussed above, subsequent secular stellar mass growth, in effect, dilutes the merger's contribution to the bulge mass. We define the ex-situ mass fraction as $M_{\rm exsitu}/M_\star$, where $M_{\rm exsitu}$ is the total stellar mass accreted from other galaxies, calculated from each galaxy's merger tree. Figure 2 shows the mean ex-situ mass fraction as a function of the B/T ratio. We find that, similar to Figure 1, bulge-less galaxies host very low ex-situ mass fractions – less than 15 per cent, on average. While a fraction of bulge-less galaxies have experienced major mergers probabilities for merger activity, with bulge-less galaxies indeed showing comparatively little merger activity over cosmic time.

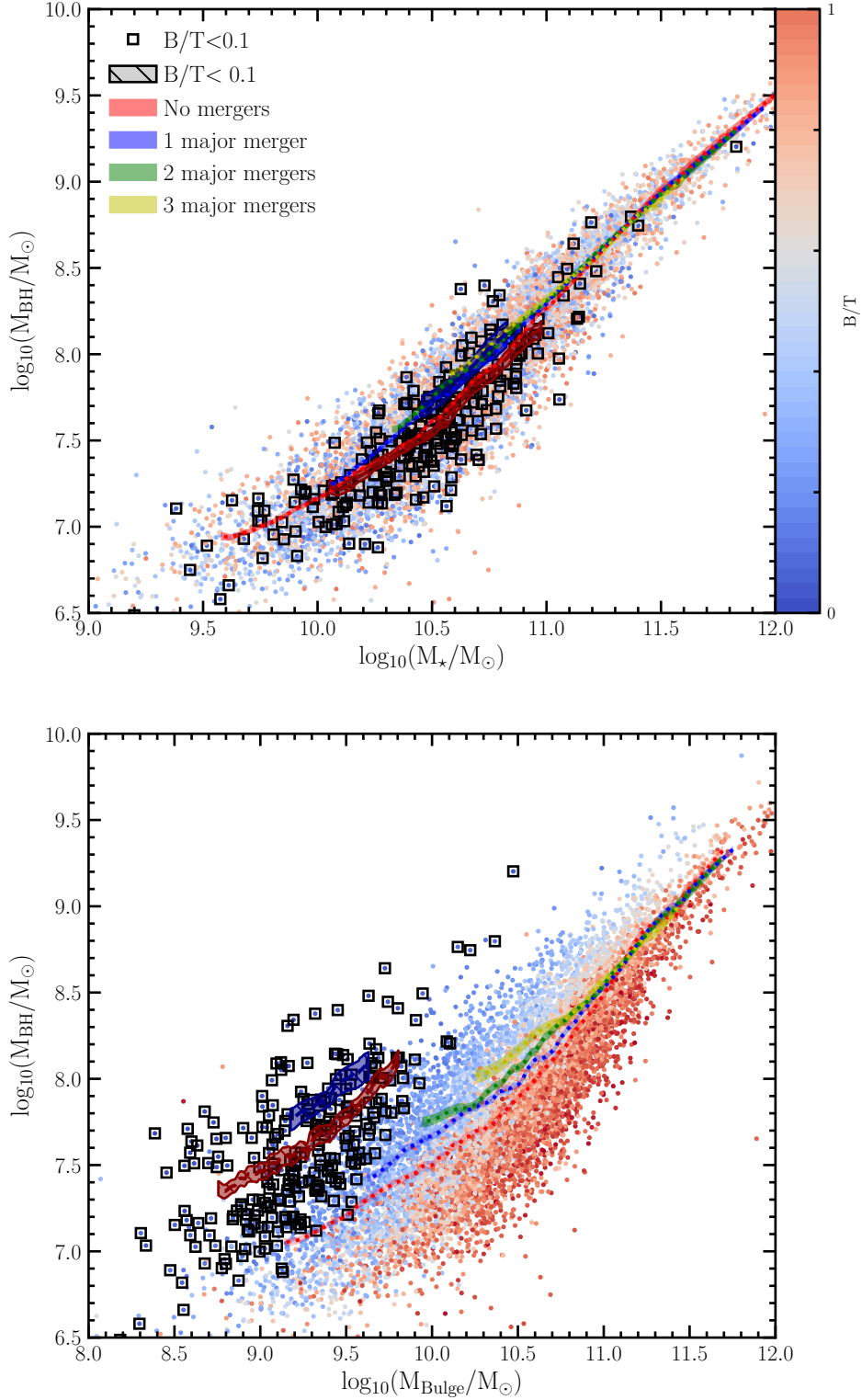

**Figure 3. Top panel:** $M_{\rm BH}$ vs $M_\star$ relation from Horizon-AGN for local massive galaxies. Galaxies are indicated by points and bulge-less galaxies are indicated by squares. Dotted coloured lines show a running mean for galaxies that have undergone 0, 1, 2 and 3 major mergers since $z = 3$ (see legend for colour coding), where the region around the line indicates the standard error on the mean. Darker coloured dashed lines with hatched regions indicate the bulge-less population only. **Bottom panel:** Same as the top panel but now with bulge mass on the *x*-axis instead of total stellar mass.

## 4 BH GROWTH OVER COSMIC TIME

### 4.1 Is there a correlation between BH growth and merger history?

We proceed by studying the $M_{BH}$–$M_{Bulge}$ and $M_{BH}$–$M_\star$ correlations in the local Universe. If mergers are primarily responsible for feeding BHs, we would expect the population of bulge-less galaxies to fall on the same $M_{BH}$–$M_{Bulge}$ and $M_{BH}$–$M_\star$ relations as the rest of the galaxy population. This is because, in the absence of mergers, the bulge-less population would have both small bulges and small BHs. Conversely, if BH feeding was preferentially produced by secular processes and accretion onto the host galaxy and *not* by galaxy mergers, then we would expect the bulge-less systems to lie on $M_{BH}$–$M_\star$ populated by the general galaxy population, but to be offset from the main $M_{BH}$–$M_{Bulge}$ locus. This is because, while secular processes steadily build their BH and stellar mass over cosmic time, their bulges will be under-massive due to the lack of major mergers. Here, we use our full sample of galaxies down to $10^9 M_\odot$, so there is some incompleteness in terms of detecting high redshift mergers towards the low mass end. However, the majority of our galaxy sample have stellar masses of $10^{10} M_\odot$ and greater, and are therefore almost entirely complete in their merger histories.

The top panel of Figure 3 shows the BH mass as a function of the total stellar mass of the galaxy ($M_\star$). Solid coloured lines show a running mean for galaxies that have undergone 0, 1, 2 and 3 major mergers since $z = 3$ (see legend for colour coding), where the width of the line indicates the standard error on the mean. Hatched regions indicate the same for the bulge-less population only. The general galaxy population is shown using the coloured dots, with the colours indicating the B/T of the galaxy in question.

The top panel of Figure 3 indicates that the number of major mergers that a galaxy has undergone does not significantly alter its position on the main locus of the $M_{BH}$–$M_\star$ correlation (offsets are visible for low-mass galaxies but these are small, < 0.1 dex per major merger). Additionally, the hatched lines, which indicate the bulge-less population, are completely consistent with the main locus of the correlation. This is evidence that mergers are not the principal driver of BH feeding, since, if that were the case, galaxies with a larger number of mergers would exhibit relatively over-massive BHs and be offset from the main locus. The bottom panel of Figure 3 shows the corresponding plot for the $M_{BH}$–$M_{Bulge}$ correlation. In this plot, the population of bulge-less galaxies lies offset above the locus traced by the general population, driven by the fact that these galaxies have under-massive bulges (due to a smaller number of mergers). As the colour coding of the points shows, in general, galaxies with lower bulge masses tend to have higher BH masses and bulge-less galaxies simply represent the tail of this trend. The fact that this trend is not present in the $M_{BH}$–$M_\star$ is strong evidence that processes that grow the bulge are not also responsible for BH growth.

Figure 4 shows the same galaxies as in Figure 3, now with each point colour-coded by the number of mergers the galaxy has undergone, using the same colour scheme as the lines in Figure 3. The bulge-less population is indicated by squares. Over-plotted are the sample of bulge-less galaxies from Simmons et al. (2013) and the sample of disc dominated galaxies from Simmons et al. (2017), with linear best fits to the observed (solid line) and simulated (dashed line) galaxies. The fits to the observed sample of at high redshift, continued stellar mass growth significantly diminishes the contribution of these events to the final mass of the galaxy at the present-day.

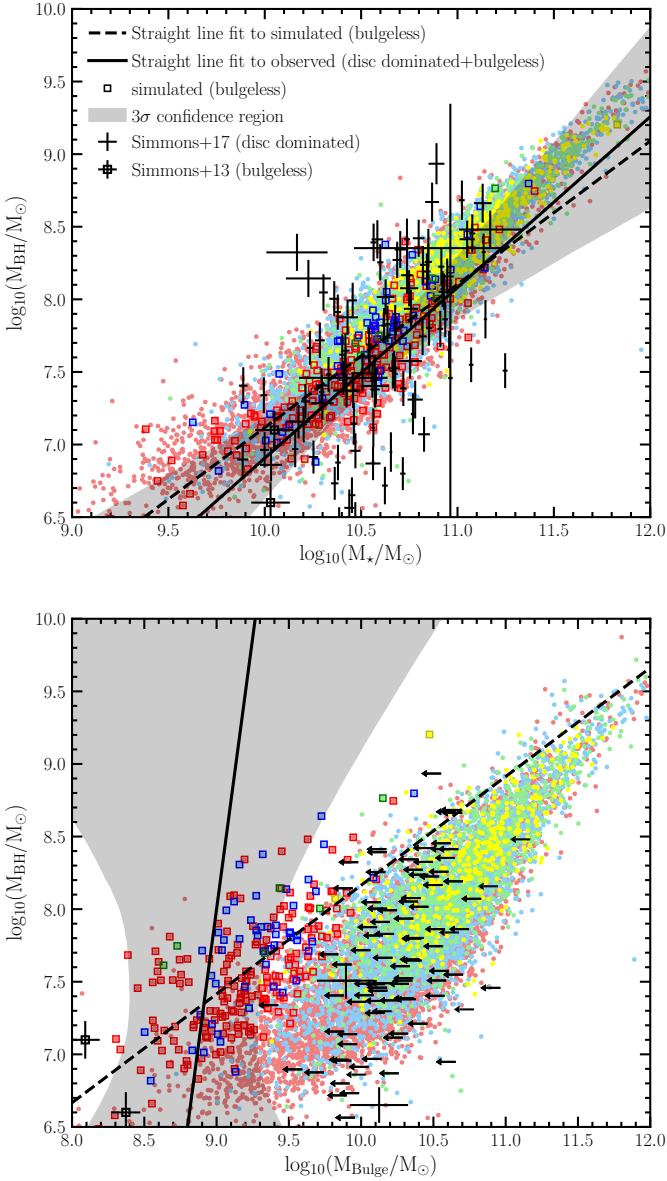

**Figure 4.** *Top panel:* $M_{BH}$ vs $M_\star$ relation for local, massive galaxies in Horizon-AGN, with points colour coded by the number of mergers as in Figure 3. Square symbols indicate bulge-less (B/T < 0.1) galaxies. Black symbols indicate observational data from Simmons et al. (2013, 2017). Dashed and solid lines show linear fits to the simulated and observed data points respectively, with the grey filled region indicating the $3\sigma$ confidence region from the fit to the observed points (Simmons et al. 2017). *Bottom panel:* Same as the top panel with bulge mass on the *x*-axis instead of total stellar mass. The arrows represent upper limits on the bulge mass. The fit properly incorporates the bulge-mass upper limits as censored data, which results in a large confidence region due to the large uncertainty on the bulge mass of these galaxies (Simmons et al. 2017).

Overall, the assumption that bulge-less galaxies have not undergone significant major-merger activity at recent ($z < 2$) epochs is robust. Progressively lower B/T ratios show rapidly diminishing disc-dominated and bulge-less galaxies are a linear regression performed by Simmons et al. (2017), incorporating errors and limits in both dimensions. The fit to the points in the bottom panel properly incorporates the bulge mass upper limits as censored data, which results in a large confidence region, due to the large uncertainty on the bulge mass of these galaxies (see Simmons et al. 2017, for more details).

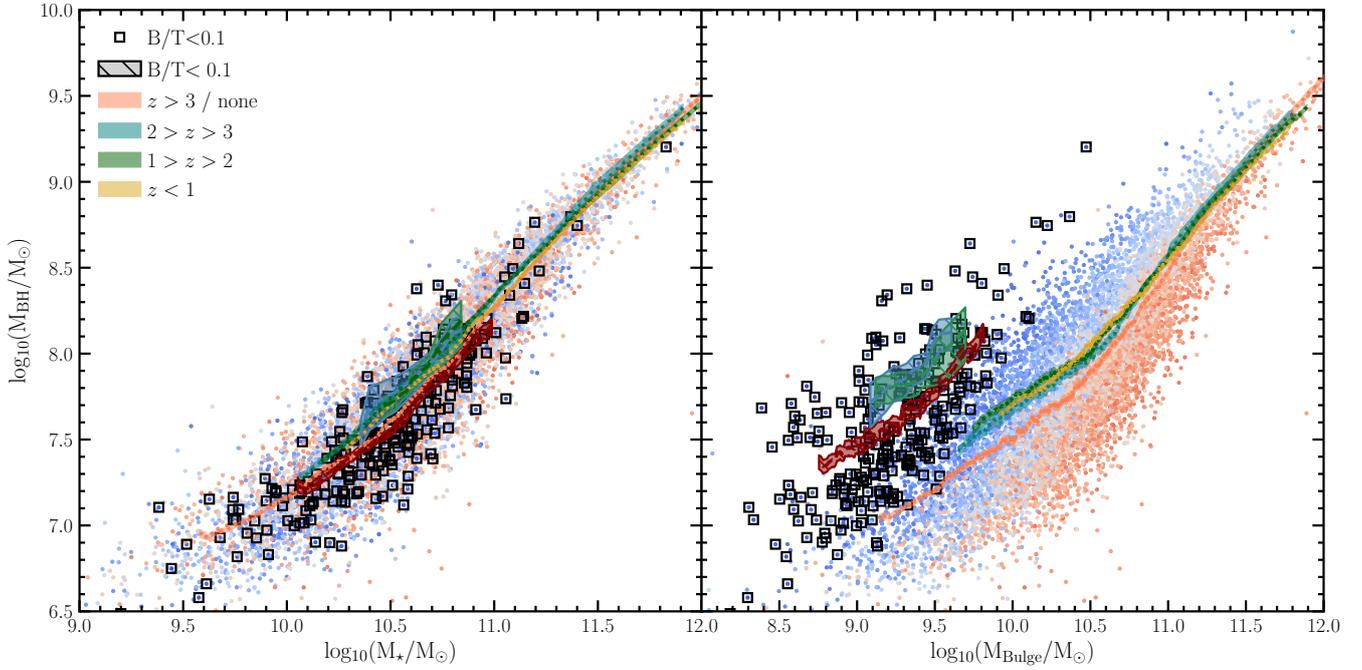

**Figure 5.** $M_{BH}$ vs $M_\star$ (left) and $M_{BH}$ vs $M_{Bulge}$ (right) relations for local massive galaxies, with the dotted coloured lines showing a running mean for galaxies that underwent their last major merger before $z = 3$, between $z = 3$ and $z = 2$, between $z = 2$ and $z = 1$ and after $z = 1$ (see legend for colour coding). The widths of the lines indicate the standard errors on the mean. Darker coloured dashed lines with hatched regions indicate the same for bulge-less galaxies.

As the top panel in this figure shows, the linear fit to the simulated bulge-less galaxies matches the slope and normalisation of the fit to the observed data, within the parameter space defined by Simmons et al. (2017). In the bottom panel, both simulated and observed bulge-less or disk-dominated galaxies lie above the $M_{Bulge}$–$M_{BH}$ relation. Although the slope of the fit to the simulated datapoints does not exactly match that of the observed data, both describe the same qualitative picture. Note that the bulk of the observed bulge masses in the bottom panel are limits. The simulated galaxies are consistent with those limits, and 3 out of 4 of the observed bulge-less galaxies which have precise measurements lie along the locus defined by the simulated bulge-less galaxies. Additionally, the majority of the simulated data points lie within the parameter space defined by the observed points.

Figure 5 again shows these two correlations for local simulated galaxies, but this time indicates how the positions of galaxies may vary, given the redshift at which their *last* major merger took place. In a similar vein to Figure 3, we find that the position of local galaxies remains largely unchanged in either correlation, irrespective of when they had their last major-merger event. Indeed, galaxies that have had major mergers around the epoch of peak cosmic star formation ($2 < z < 3$) do not deviate from the main locus of the correlation, indicating that the gas richness of these major mergers have little impact on the overall growth of their BHs.

Our analysis so far has focussed on galaxies in the local Universe and has shown that mergers are unimportant in terms of the *cumulative* evolution of BHs over cosmic time. It is also instructive to study whether merger activity might have a transient impact on the $M_{BH}$–$M_{Bulge}$ and $M_{BH}$–$M_\star$ correlations at high redshift. We complete our analysis by studying the redshift evolution of these correlations, and exploring whether the impact of major mergers may be higher in the high-redshift Universe. In Figures 6 and 7, we show the redshift evolution of these correlations in the simulation, with mean locii indicated for galaxies that have had 0, 1, 2 and 3 major mergers before the redshift in question ($z = 0$, 0.5 and 2.5; which correspond to look-back times of 0, 5 and 11 Gyrs respectively) shown using the coloured lines. The colour coding is the same as that used in Figure 3.

This figure shows that the number of major mergers a galaxy experiences does not alter its position on the *evolving* correlations as a function of redshift. Indeed, if major mergers were the principal driver of BH growth, then galaxies would be expected to show large offsets from the mean locus (which would induce a large scatter), before enough merging has taken place to put them on the relation at the present day. However, Figure 6 indicates a persistently tight correlation, as these relations build up steadily over cosmic time, the opposite to what would be expected if BH growth were episodic and driven by largely stochastic events like major mergers. *Thus, major-merger activity of any kind is unlikely to be driving significant BH growth at any epoch.*

Our analysis suggests that whatever processes dominate the *overall* stellar-mass growth of the galaxy population over cosmic time, also drive the growth of their constituent BHs. Furthermore, BH mass does not correlate as well with the part of the galaxy, i.e. the bulge, that is preferentially built in mergers. Together, this indicates that BH growth tends to occur largely by secular means, without recourse to mergers.

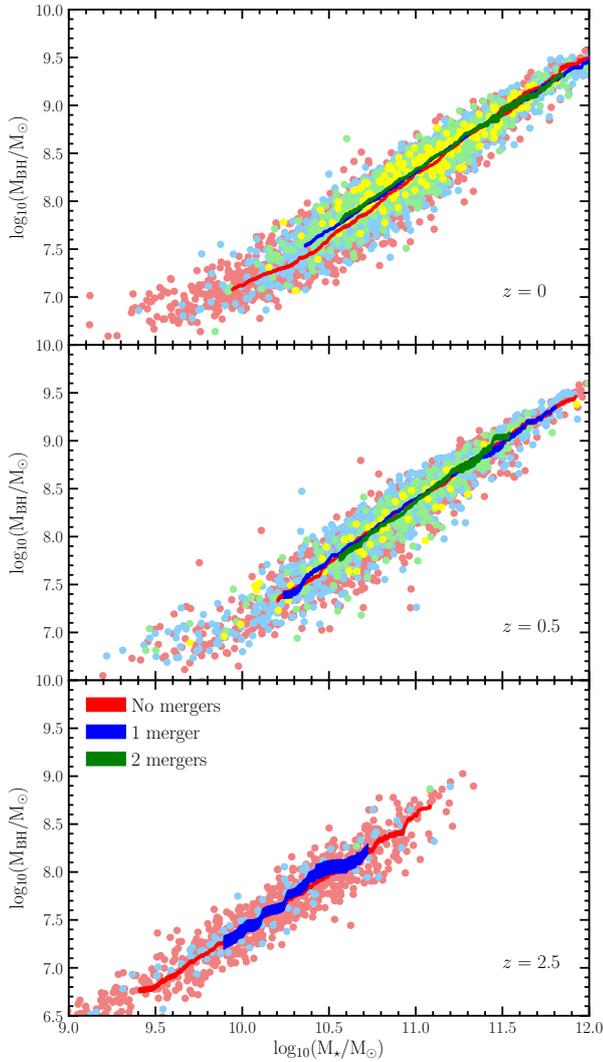

**Figure 6.** Evolution of the $M_{BH}$–$M_\star$ relation in Horizon-AGN for local massive galaxies. Solid coloured lines show a running mean for galaxies that have undergone 0 (red), 1 (blue) and 2 (green) major mergers before the redshift indicated in each panel, where the width of the line indicates the standard error on the mean.

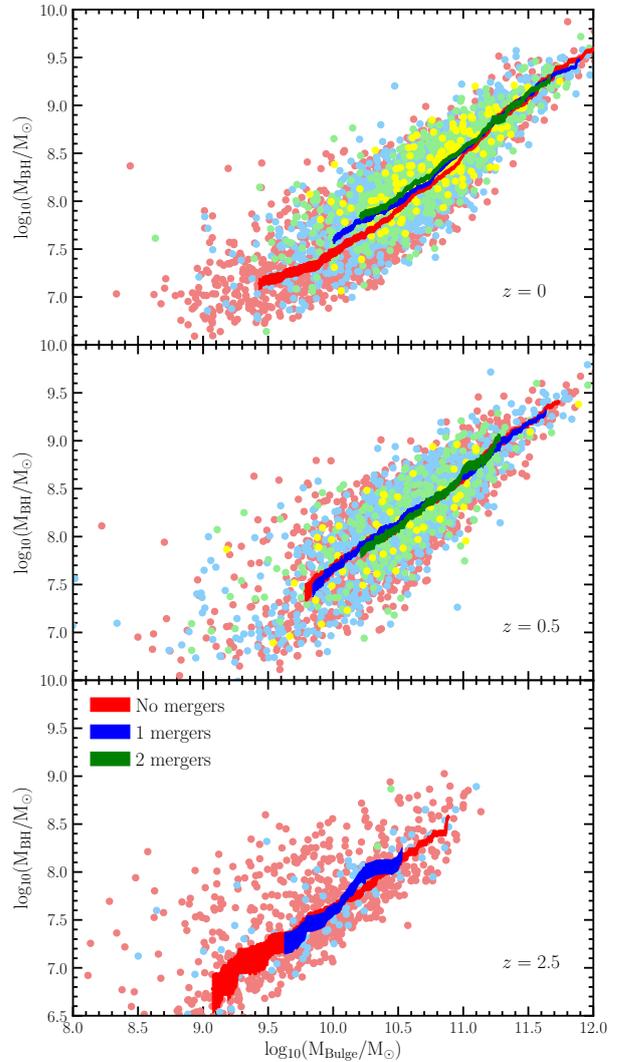

**Figure 7.** Evolution of the $M_{BH}$–$M_{Bulge}$ relation in Horizon-AGN. As in Figure 6, solid coloured lines show a running mean for galaxies that have undergone 0 (red), 1 (blue) and 2 (green) major mergers before the redshift indicated in each panel, where the width of the line indicates the standard error on the mean.

### 4.2 Contribution of mergers to the cosmic BH accretion budget

So far, we have shown that galaxies that have undergone mergers do not lie on a different $M_{BH}$ vs $M_\star$ relation to their non-merging counterparts. While this is evidence that BH growth does not preferentially take place in mergers, it is useful to precisely quantify the fraction of the cosmic BH accretion budget which is directly attributable to the merger process over cosmic time. Recent work, that has studied the proportion of the star formation budget that is directly driven by major and minor mergers (Martin et al. 2017), has shown that only 25 per cent of the stellar mass in today's Universe is directly triggered by merging, with major and minor mergers accounting for 10 and 15 per cent of this value respectively. Here, we perform a corresponding study of BH growth and quantify the proportion of the BH accretion budget that is attributable to major and minor mergers.

We perform our analysis by tracking the mass evolution of each of the BHs hosted by one of our galaxies at $z = 0$. In a similar vein to Martin et al. (2017) who studied merger-driven star formation activity, we first define a merger-driven enhancement of the BH accretion rate, $\xi$, as the ratio of the mean specific BH accretion rate in the merging galaxies to that in their non-merging counterparts:

$$\xi(M_{BH}, z) = \frac{\langle \dot{M}_{BH}/M_{BH}(M_{BH}, z)\rangle_m}{\langle \dot{M}_{BH}/M_{BH}(M_{BH}, z)\rangle_{non}}, \quad (3)$$

where $\dot{M}_{BH}$ is the BH accretion rate. $\langle \dot{M}_{BH}/M_{BH}(M_{BH}, z)\rangle_m$ is the mean specific accretion rate for the merging population at a given redshift, $z$, and $\langle \dot{M}_{BH}/M_{BH}(M_{BH}, z)\rangle_{non}$ is the same for galaxies that are not merging. Galaxies are defined as merging if they have had undergone a merger (major or minor) within the last Gyr or will undergo a merger in the next Gyr. Our results are robust to changes in this timescale: doubling or halving this number changes the contribution of mergers to the cosmic star formation budget by less that 5 per cent.

We use this enhancement to estimate the fraction of BH accre-tion that would have occurred in the merger progenitors anyway, had they not been in the process of merging. For example, if $\xi$ is a factor of 2 then, on average, around half the BH accretion in the merging system in question is likely driven by other processes (see e.g. Kaviraj et al. 2013b; Lofthouse et al. 2017; Martin et al. 2017, for a similar discussion of star formation activity). By subtracting the BH accretion that would have occurred anyway, had the merger not taken place, we can then measure the fraction of BH accretion that is *directly triggered* by mergers ($f$) as follows:

$$f = \frac{m_{new,m}(M_{BH}, z)\left[1 - 1/\xi(M_{BH}, z)\right]}{m_{new,total}(M_{BH}, z)}, \quad (4)$$

where $m_{new,m}(M_{BH}, z)$ is the total mass accreted onto BHs in merging systems in a given BH mass and redshift bin and $m_{new,total}(M_{BH}, z)$ is the total mass accreted onto BHs in the stellar mass and redshift bin in question[1]. To ensure that our sample is complete down to a merger mass ratio of 1:10, we restrict ourselves to galaxies with stellar masses $M_\star > 10^{9.5} M_\odot$ at all redshifts.

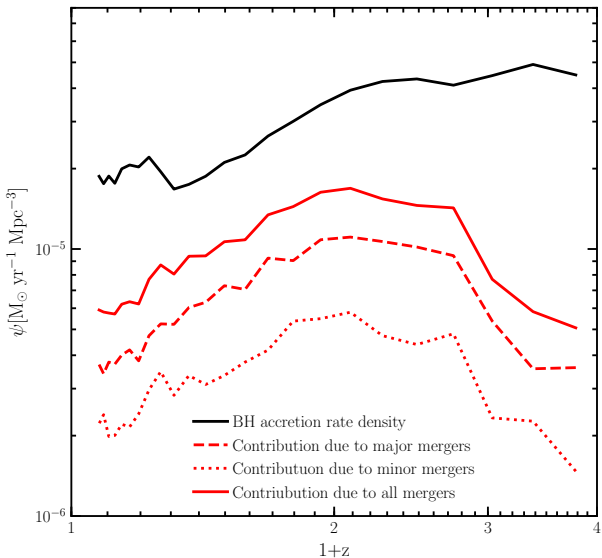

**Figure 8.** The BH accretion rate density for galaxies with $M_\star > 10^{9.5} M_\odot$ as a function of redshift from Horizon-AGN (black). The red lines indicate the portion of the BH accretion rate density that is a direct result of major (dashed line), minor (dotted line) and major + minor (all) mergers (solid line). The small jump in accretion at low redshift corresponds to the introduction of an additional grid refinement level at $z = 0.26$.

Figure 8 shows the cosmic BH accretion rate density in Horizon-AGN as a function of redshift. The BH accretion rate density decreases with redshift. The contribution due to mergers increases towards $z = 1$, peaking at around $z = 1.1$, and decreasing towards the present day. At all times, major mergers outweigh the contribution of minor mergers to the BH accretion rate density, even though minor mergers account for the majority of galaxy interactions (e.g. Kaviraj et al. 2015b; Lotz et al. 2011).

Since black hole accretion in Horizon-AGN is modelled using the Bondi-Hoyle-Lyttleton rate (Equation 1), the increase in gas density around the BH corresponds directly to an increase in the accretion rate. Mergers are least significant at high redshift, where galaxies already host high densities of gas (e.g. Geach et al. 2011), which enables efficient BH growth through the secular accretion of low angular momentum gas over short timescales (Dubois et al. 2012).

The small jump in accretion rate density observed at low redshift is due to the implementation of an additional AMR grid refinement at $z = 0.26$. This increases the local density in gas cells, thus increasing the accretion rate onto the black hole. Mass accretion after $z = 0.26$ only accounts for 12 per cent of total mass accreted by black holes since $z = 3$, so the effect of grid refinement does not alter our qualitative conclusions.

Figure 9 shows the cumulative fraction of BH mass (in galaxies more massive than $10^{9.5} M_\odot$) that is triggered by major and minor mergers as a function of redshift. At the present day, only ∼35 per cent of the BH mass in massive galaxies is directly attributable to the merger process – of this ∼22 per cent is driven by major mergers while the rest (∼13 per cent) is driven by minor mergers. Mergers are, therefore, minority contributors to the BH accretion budget over cosmic time. It is worth noting that these values are not a strong function of galaxy mass. The fraction of BH mass that is created as a direct result of mergers increases from ∼ 25 per cent in galaxies with stellar masses around $10^{10} M_\odot$ to ∼ 40 per cent in galaxies with stellar masses of $10^{11.5} M_\odot$ or greater. However, across the range of stellar masses considered in this study, the majority of the BH mass is created via secular processes, not mergers.

Finally, we note that, while only ∼ 25 per cent of black hole growth globally is the direct result of major mergers, a small fraction of galaxies do grow most of their BH mass during major mergers. 28 per cent of galaxies that have undergone at least one major merger since $z = 3$ have more than half of their total black hole mass built up as a direct result of major mergers during this time; this number is reduced to just 12 per cent when all galaxies are taken into account. The fact that the BH growth of a small fraction of galaxies is dominated by merging is likely responsible for the small increase in scatter towards higher redshift indicated by Figure 7.

## 5 SUMMARY

A consistent picture is now emerging of the role that galaxy mergers play in driving stellar mass and BH growth across cosmic time, and particularly in the early Universe. Both theoretical and observational work now indicates that major mergers (and mergers in general) do not enhance star-formation activity around the epoch of peak cosmic star formation (e.g. Lofthouse et al. 2017; Fensch et al. 2017). In other words, the bulk of the star formation that takes place at these epochs is driven secularly via cosmological accretion and not triggered by merging. And since the bulk of the stellar mass in today's galaxies forms around this epoch, the majority of today's stellar mass (∼75 per cent, see Martin et al. 2017) is also unrelated to merging.

This particular study has used Horizon-AGN, a cosmological hydrodynamical simulation, to extend this analysis to BH growth. Our results indicate that a similar picture to that for star formation activity likely holds for accretion on to BHs. The majority (∼65 per cent) of the cumulative BH growth in today's massive galaxies takes place via secular processes, with the remaining ∼35 per cent attributable to either major or minor mergers. Our key findings can be summarised as follows:

• *Almost all bulge-less galaxies have undergone no major mergers since $z = 1$. However, ∼25 per cent of such systems*

---

[1] Equation 4 above is the BH-accretion equivalent of Equation 2 in Martin et al. (2017).

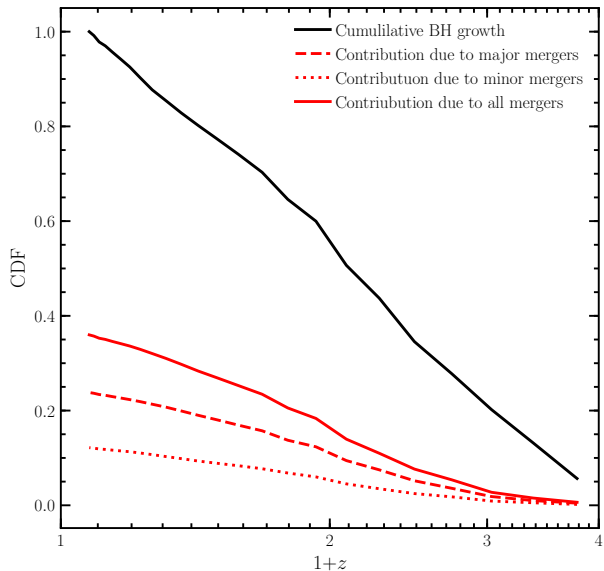

**Figure 9.** The cumulative fraction of BH mass in today's massive galaxies that has already been assembled as a function of redshift (black line). The contribution from major mergers, minor mergers and major + minor (all) mergers are shown using the dashed, dotted and solid red lines respectively. Only ∼35 per cent of the BH mass in massive galaxies at the present day is directly attributable to merger activity. ∼22 per cent is driven by major mergers and ∼13 per cent is driven by minor mergers. The bulk (∼ 65 per cent) of the BH mass build-up over cosmic time is unrelated to merging and is therefore be a result of secular processes.

have had a major merger since $z = 3$ (although, on average, more than 85 per cent of their stellar mass at $z = 0$ is formed in-situ), indicating that disk rebuilding in gas-rich mergers may play a role in building these systems. However, the assumption made in many observational studies, that bulge-less galaxies have undergone no major mergers over most of cosmic time, is typically robust.

• *Bulge-less galaxies lie on the same $M_{BH}$–$M_\star$ relation as the general galaxy population.* The number of major (mass ratios greater than 1 : 4) or minor mergers (mass ratios between 1 : 4 and 1 : 10) that a galaxy has undergone does not alter a galaxy's position on the $M_{BH}$–$M_\star$ relation, indicating that mergers are not a significant mechanism for feeding the BH.

• *Bulge-less galaxies lie offset from the $M_{BH}$–$M_{Bulge}$ relation observed in the general population.* This relation is not as tight as the $M_{BH}$–$M_\star$ relation, with the number of mergers having a larger effect on the position of a galaxy on the $M_{BH}$–$M_{Bulge}$ relation than on the $M_{BH}$–$M_\star$ relation. The offset of the bulge-less galaxies is driven by the fact that these galaxies have normal black holes but under-massive bulges (due to a smaller number of mergers).

• *Mergers are directly responsible for a minority of BH growth over cosmic time.* Only ∼35 per cent of the BH mass in galaxies more massive than $10^{9.5} M_\odot$ in today's Universe is directly attributable to mergers. ∼22 per cent is driven by major mergers and ∼13 per cent is driven by minor mergers. Secular processes, therefore, account for the creation of the majority (∼65 per cent) of BH mass over the lifetime of the Universe.

## ACKNOWLEDGEMENTS

We thank the referee, Smita Mathur, for many constructive comments that improved this paper. GM acknowledges support from the Science and Technology Facilities Council [ST/N504105/1]. SK acknowledges a Senior Research Fellowship from Worcester College Oxford. MV acknowledges funding from the European Research Council under the European Community's Seventh Framework Programme (FP7/2007-2013 Grant Agreement no. 614199, project 'BLACK'). JD acknowledges funding support from Adrian Beecroft, the Oxford Martin School and the STFC. BDS acknowledges support from the National Aeronautics and Space Administration (NASA) through Einstein Postdoctoral Fellowship Award Number PF5-160143 issued by the Chandra X-ray Observatory Center, which is operated by the Smithsonian Astrophysical Observatory for and on behalf of NASA under contract NAS8-03060. RJS gratefully acknowledges funding from the Ogden Trust. This research has used the DiRAC facility, jointly funded by the STFC and the Large Facilities Capital Fund of BIS, and has been partially supported by grant Spin(e) ANR-13-BS05-0005 of the French ANR. This work was granted access to the HPC resources of CINES under the allocations 2013047012, 2014047012 and 2015047012 made by GENCI and is part of the Horizon-UK project. Marc Sarzi and Donna Rodgers-Lee are thanked for useful comments.

This paper has been typeset from a TEX/LATEX file prepared by the author.